\documentclass[11pt]{article}
\pdfoutput=1
\usepackage{amsmath,amssymb,color,epsfig,cite}
\usepackage{graphicx}
\usepackage{subfigure}
\usepackage{setspace}

\usepackage{amsfonts}

\newcommand{\hoch}[1]{$\, ^{#1}$}

\makeatletter
\@addtoreset{equation}{section}
\makeatother

\newcommand{\be}{\begin{equation}}
\newcommand{\ee}{\end{equation}}
\newcommand{\bea}{\setlength\arraycolsep{2pt} \begin{eqnarray}}
\newcommand{\eea}{\end{eqnarray}}
\newcommand{\nn}{\nonumber}

\def\ft#1#2{{\textstyle{\frac{\scriptstyle #1}{\scriptstyle #2} } }}
\def\fft#1#2{{\frac{#1}{#2}}}

\def\0{{\sst{(0)}}}
\def\1{{\sst{(1)}}}
\def\2{{\sst{(2)}}}
\def\3{{\sst{(3)}}}
\def\4{{\sst{(4)}}}
\def\5{{\sst{(5)}}}
\def\6{{\sst{(6)}}}
\def\7{{\sst{(7)}}}
\def\8{{\sst{(8)}}}
\def\sst#1{{\scriptscriptstyle #1}}

\begin{document}

\begin{center}
{\large {\bf Action Growth of Dyonic Black Holes and Electromagnetic Duality}}

\vspace{10pt}
Hai-Shan Liu\hoch{1,2} and H. L\"u\hoch{2}

\vspace{10pt}

\hoch{1}{\it Institute for Advanced Physics \& Mathematics,\\
Zhejiang University of Technology, Hangzhou 310023, China}

\vspace{10pt}

\hoch{2}{\it Center for Joint Quantum Studies, Tianjin University, Tianjin 300350, China}

\vspace{30pt}

\underline{ABSTRACT}

\end{center}

Electromagnetic duality of Maxwell theory is a symmetry of equations but not of the action. The usual application of the ``complexity=action'' conjecture would thus loose this duality.  It was recently proposed in arxiv:1901.00014 that the duality can be restored by adding some appropriate boundary term, at the price of introducing the mixed boundary condition in the variation principle. We present universal such a term in both first-order and second-order formalism for a general theory of a minimally-coupled Maxwell field. The first-order formalism has the advantage that the variation principle involves only the Dirichlet boundary condition. Including this term, we compute the on-shell actions in the Wheeler-De Witt patch and find that the duality persists in these actions for a variety of theories, including Einstein-Maxwell, Einstein-Maxwell-Dilaton, Einstein-Born-Infeld and Einstein-Horndeski-Maxwell theories.

\vfill {\footnotesize Emails: hsliu.zju@gmail.com \, mrhonglu@gmail.com}

\thispagestyle{empty}

\pagebreak

\tableofcontents
\addtocontents{toc}{\protect\setcounter{tocdepth}{2}}



\section{Introduction}

The AdS/CFT correspondence builds a bridge between classical gravity in asymptotic anti-de Sitter (AdS) spacetime and some strongly-coupled conformal field theory (CFT) living on its boundary \cite{adscft1,adscft2,adscft3,adscft4}. It provides a new tool for studying the dynamics of various strongly coupled quantum field theories \cite{xxt1,xxt2,xxt3,xxt4}. Recently, holographic connections between the quantum complexity of a state on the boundary and some physical quantities in the bulk theory have attracted considerable attention. Conjectures such as ``complexity = volume" (CV) \cite{cv1,cv2} and ``complexity = action" (CA) \cite{ca1,ca2} have been proposed. Many works have been done to widely explore the properties of the holographic complexity related to these two conjectures \cite{camyers,cav2,cav3,cav4,cav5, Jiang:2018sqj,cav7,cav8,cav9,cav10,cav11,cav12,cav13,ldbd, twoh1, ptwoh, masgr, frcg, fr, frmas, lovelock,oneh,kim,feng} and further to generalize these two conjectures \cite{cav1,cav6,jiang,cvt1,sub0,Fan:2019mbp}.

The CA conjecture states that the complexity of the boundary state is related to the on-shell classical gravitational action evaluated on the so-called ``Wheeler-DeWitt (WDW) patch'' of the asymptotically-AdS black hole, namely
\be
{\cal C} = \fft{I_{\rm WDW}}{\pi} \,.
\ee
The WDW patch is a spacetime area enclosed by light rays, as illustrated in Fig.~1.

\begin{figure}[htp]
\begin{center}
\includegraphics[width=200pt]{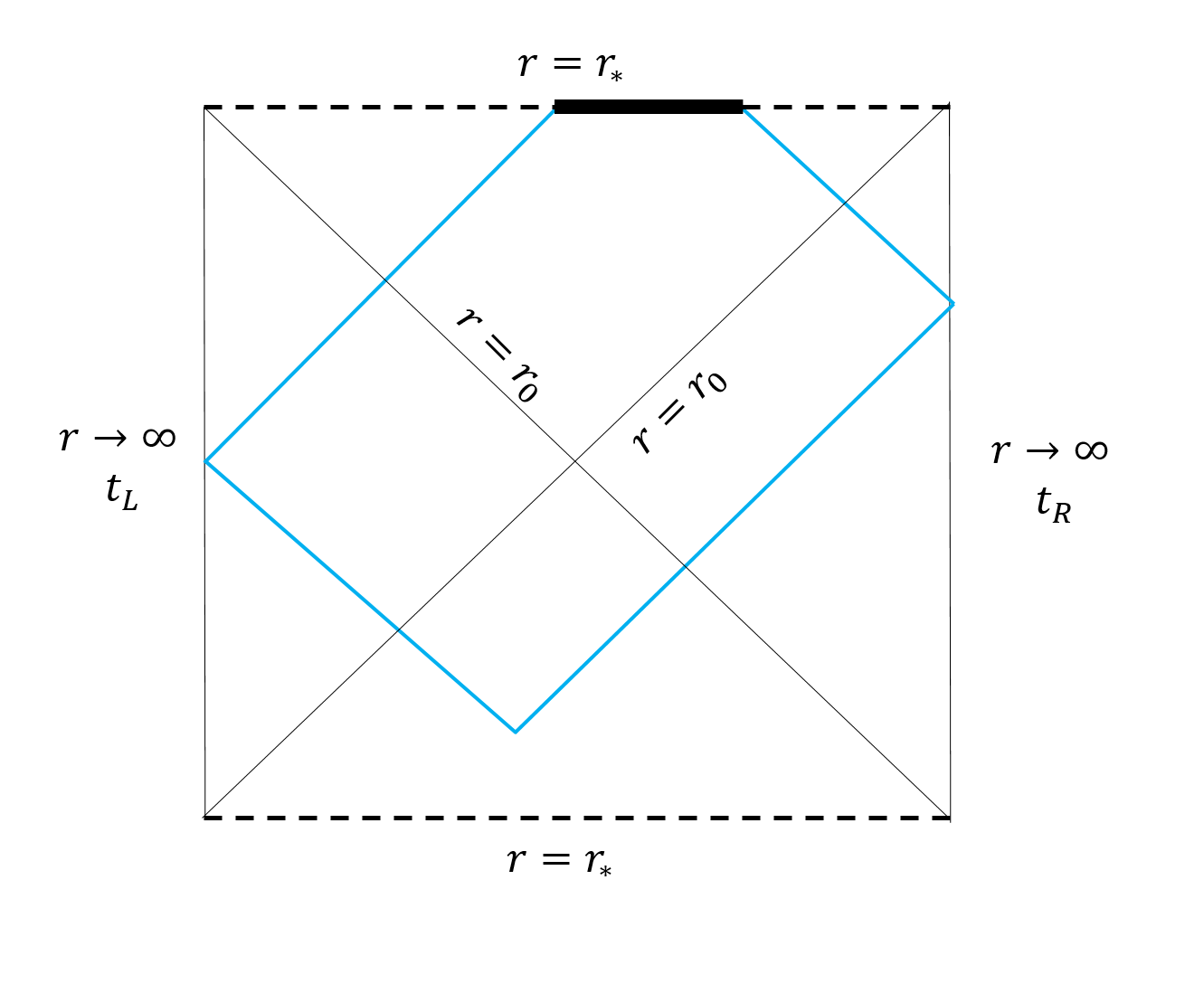}
\includegraphics[width=200pt]{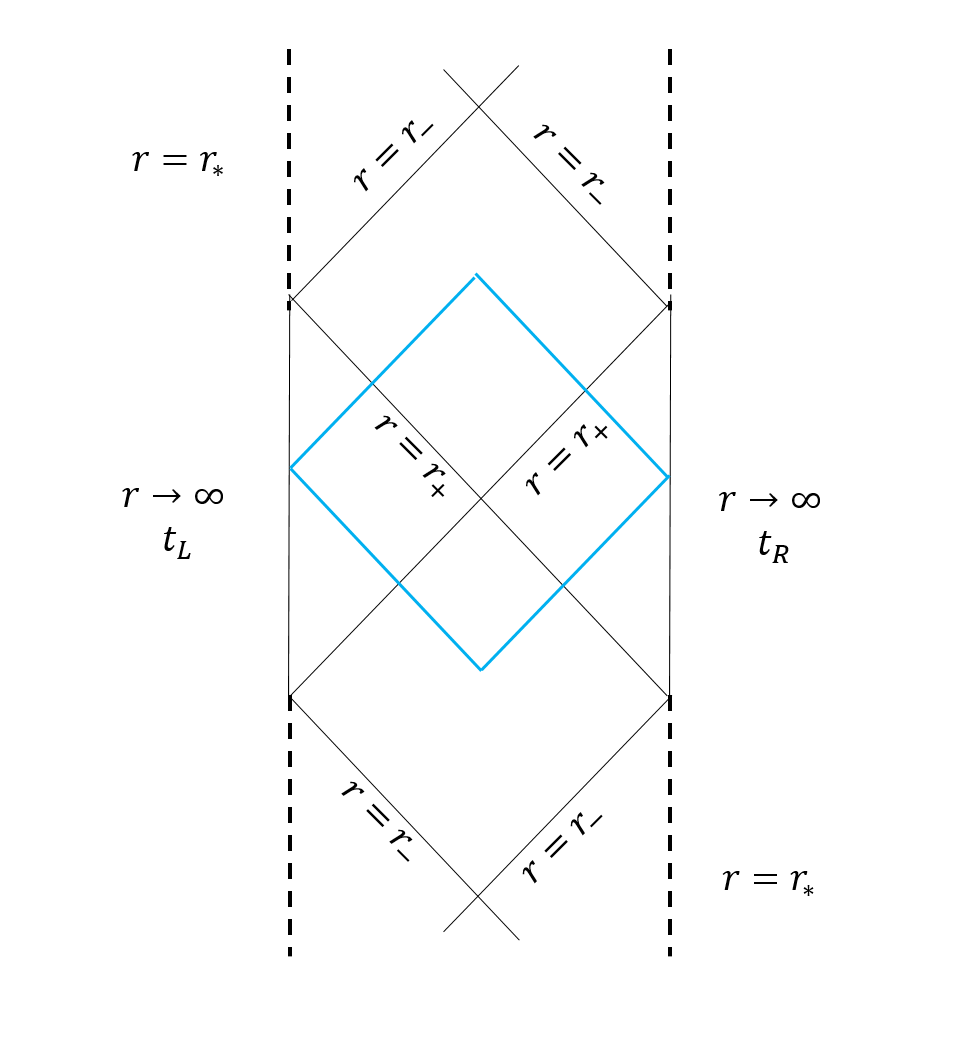}
\end{center}
\caption{\setstretch{1.0}\small\it Plots of Wheeler-DeWit patches.  The left panel shows the causal structure of Schwardzchild-like black hole which has only one event horizon located at $r_0$, the region enclosed by blue lines is the WDW patch. The right panel shows the causal structure of Reissner-Nordstrorm like black hole which has two horizons located at $r_+$ and $r_-$, and the WDW patch is the area enclosed by the blue lines.  } \label{figure1}
\end{figure}

It was well known that in gravities or supergravities, many global symmetries can be realized only at the level of equations, but not directly in the action. Indeed the well-known Cremmer-Julia $E_{7(+7)}$ global symmetry of maximum supergravity in four dimensions is a symmetry of equations. The simplest example to illustrate this is perhaps the Maxwell theory in general curved spacetime with the Lorentzian signature.  The action is
\be
I=\fft{\kappa}{16\pi} \int_M d^4x \sqrt{-g} (-\ft14 F^2)\,,\qquad F=dA\,.
\ee
The equation of motion and Bianchi identity, i.e.
\be
d{*F}=0\,,\qquad dF=0\,,
\ee
can be interchanged by interchanging $F$ and $*F$, giving rise to the electromagnetic duality. However, this symmetry cannot be realized in the action for two reasons.  The first is that the fundamental field in Lagrangian is the gauge potential $A$ rather than $F$; the second is that in Lorentzian signature, we have the identity
\be
(*F)^2 = - F^2\,.
\ee
Indeed the electric and magnetic charges $(Q_e,Q_m)$ contribute symmetrically to the energy-momentum tensor, e.g.~$Q_e^2 + Q_m^2$, but they contribute anti-symmetrically to the on-shell action, e.g.~$Q_e^2-Q_m^2$. It follows that the standard application of the CA conjecture would loose the  duality in the Einstein-Maxwell theory. This disappearance of the electromagnetic duality in holographic complexity from the CA conjecture was recently singled out in \cite{susk} and \cite{goto}.

There is more than one way to restore the electromagnetic duality.  A double-field formalism can make the duality manifest even in the off-shell action at the price of introducing extra degrees of freedom that need to be removed by constraints \cite{Cremmer:1998px}. (See also \cite{Deser:1981fr,Deser:1996xu}.) For the on-shell action, the restoration of the duality is simpler and can be done simply by introducing a Maxwell boundary term \cite{goto}
\be
\fft{\kappa \gamma}{16\pi} \int_{\partial M} d\Sigma_\mu\, F^{\mu\nu} A_\nu\,,
\ee
with appropriate coefficient $\gamma$.  Unlike the Gibbons-Hawking surface term \cite{gbhk}, this boundary term was not required by the variation principle. In fact except for $\gamma=0,1$, the variation principle would have to deal with mixed Dirichlet and Neumann boundary conditions, which typically requires extra degrees of freedom on the boundary. The $\gamma=1$ case was introduced in \cite{gbhk,mbbb,mbhk,Lu:2013ura} and it has an effect of performing the Legendre transformation in black hole thermodynamics associated with the electric charge $Q_e$ and its canonical potential $\Phi_e$.

It was demonstrated for the RN-AdS black hole \cite{goto} that the on-shell action can recover the full electromagnetic duality for appropriate $\gamma$. This leads to an important question whether there is a universal $\gamma$ for restoring the electromagnetic duality in the on-shell action.
The situation for purely electric or purely magnetic AdS black holes in a family of Einstein-Maxwell-Dilaton (EMD) theory was also examined in \cite{goto}.

In this paper, we examine a variety of theories that exhibits the electromagnetic duality at the level of equations of motion. These theories all admit dyonic AdS black holes, which allow us to evaluate the on-shell action and determine whether there is a universal term that can restore the duality in the on-shell action.

We first propose in section 2 a first-order formalism for a general theory of the Maxwell field that couples to gravity minimally.  In this first-order system, we can add a total derivative term to restore the electromagnetic duality of the on-shell action. The advantage of the first-order formalism is that the variation principle involves only the Dirichlet boundary condition. We show that the on-shell action is precisely equivalent to that with the Maxwell boundary term in the second-order formalism. This yields a general formula for the required boundary term for restoring the electromagnetic duality. Since we are unable to give a mathematical proof for this formula,
in subsequent sections, we examine the on-shell action for dyonic black holes in Einstein-Maxwell, Einstein-Maxwell-Dilaton, Einstein-Born-Infeld and Einstein-Hordenski-Maxwell theories and confirm our conjecture.  In doing so, we obtain the explicit late-time action growth rate of the dyonic black holes in these theories. We conclude the paper in section 7.

\section{First-order and second-order formalisms}

In this section, we set up the general formalism for calculating the on-shell action involving the Maxwell field. We shall apply this formalism throughout this paper to evaluate the on-shell actions for various black hole solutions. We consider a general theory of the minimally-coupled Maxwell field in four-dimensional curved spacetime,  We begin with the first-order formalism and the action is given by
\be
 {\cal I }_F = \int_M d^4x \sqrt{-g} \big[ {\cal L}_F(F_{\mu\nu}) - \ft 12  \epsilon^{\mu\nu\rho\sigma} \nabla_\mu (F_{\nu\rho}) B_\sigma + \ft12\gamma \epsilon^{\mu\nu\rho\sigma} \nabla_\mu (F_{\nu\rho} B_\sigma)\big] \,.
\ee
Here the antisymmetric Maxwell field strength $F_{\mu\nu}$ is a fundamental field and $B_\mu$ is an auxiliary field. The $\gamma$-term is a total derivative of a bilinear term and gives no contribution to the equations of motion; however, it can alter the on-shell action. Variation of the action with respect to $B_\mu$ gives the Bianchi identity
\be
\nabla_{[\mu} F_{\nu\rho]} = 0 \,,\label{genbi}
\ee
which implies that $F=dA$. Variation of $F_{\mu\nu}$ gives
\be
\widetilde F^{\mu\nu} = \epsilon^{\mu\nu\rho\sigma} \nabla_\rho B_\sigma\,,\qquad\hbox{with}\qquad
\widetilde F^{\mu\nu} \equiv -2 \fft{\partial {\cal L}_F}{\partial F_{\mu\nu}}\,.
\ee
Consequently, $\widetilde F^{\mu\nu}$ satisfies the equation of motion
\be
\nabla_\mu \widetilde F^{\mu\nu}=0\,.\label{tFeom}
\ee
Applying the Bianchi identity and equation of motion, the on-shell action is
\bea
{\cal I }^F_{\rm on-shell} &=& \int_M d^4x \sqrt{-g} \big[ {\cal L}_F + \ft12\gamma \epsilon^{\mu\nu\rho\sigma}F_{\mu\nu} \nabla_\rho B_\sigma \big]\nn\\
  &=&  \int_M d^4x \sqrt{-g} \big[ {\cal L}_F +\ft12\gamma \widetilde F^{\mu\nu} F_{\mu\nu} \big] \,. \label{actF}
\eea
It should be emphasized that in this first-order formalism, the standard Dirichlet boundary condition should be adopted in the application of the variation principle, namely the variation of the fundamental fields $\delta F_{\mu\nu}$ and $\delta B_\mu$ both vanish on the boundary.

Alternatively, we consider the second formalism where $F = dA$ and $A$ is the Maxwell potential.  In this case, the Bianchi identity is automatic. The action is
\be
{\cal I }_A = \int_M d^4x \sqrt{-g} {\cal L}_F (F_{\mu\nu})  + \int_{\partial M} d\Sigma_\mu S^\mu\,,
\ee
where $d\Sigma_\mu$ is the infinitesimal boundary volume element pointing to the normal direction of the boundary, and  $S^\mu(A)$ is some appropriate boundary term that is to be determined.  The variation with respect to $A_\nu$ gives (after integrating out the total derivative)
\be
\delta {\cal I}_A =
\int_M (\nabla_\mu \widetilde F^{\mu\nu}) \delta A_\nu +
\int_{\partial M} d\Sigma_\mu (-\widetilde F^{\mu\nu}\delta A_\nu + \delta S^\mu)\,.
\ee
Thus the equation of motion is (\ref{tFeom}), as in the first-order formalism.  Furthermore, the boundary structure of $\delta {\cal I}_A$ suggests that we can take
\be
S^\mu = \gamma \widetilde F^{\mu\nu} A_\nu\,.
\ee
In particular, in Maxwell theory with $\widetilde F_{\mu\nu}=F_{\mu\nu}$, setting $\gamma=1$ gives the Legendre transformation interchanging the role of
$\widetilde F^{\mu\nu}$ and $A_\nu$.  The Dirichlet boundary condition $\delta A_\mu|_{\partial M}=0$ is then switched to become the Neumann boundary condition $n^\mu\partial_{\mu} \delta A_\nu|_{\partial M}=0$.  In the first law of black hole thermodynamics, it becomes the Legendre transformation interchanging the role of the electric charge and its chemical potential. Here we let the constant $\gamma$ be arbitrary.

The on-shell action is then
\bea
{\cal I }_A &=& \int_M d^4x \sqrt{-g} {\cal L}_F + \gamma \int_{\partial M}
d\Sigma_\mu \widetilde F^{\mu\nu} A_\nu\nn\\
&=& \int_M d^4x \sqrt{-g} \left({\cal L}_F + \gamma \nabla_\mu (\widetilde F^{\mu\nu} A_\nu)\right)\,.\label{actA}
\eea
The Stock's theorem was applied in the second equality. It is clear that the on-shell actions (\ref{actF}) and (\ref{actA}) are the same. The price to pay in the second-order formalism  is that one has to adopt the mixed Dirichlet and Neumann boundary condition, which typically requires additional degrees of freedom on the boundary.  However, in terms of the evaluation of the on-shell action, the two formalisms give the identical result and it is more straightforward to use (\ref{actA}) to evaluate the action.

Thus we see that the action of the (general) Maxwell theory is not uniquely defined.  In the first-order formalism, the ambiguity amounts to a total derivative, with an arbitrary dimensionless constant $\gamma$. In the second-order formalism, it represents itself as a surface term. However, this surface term is not required by the variation principle such as the Gibbons-Hawking term that must have a specific coefficient. The effect is like a Legendre transformation in black hole thermodynamics at the price of introducing mixed boundary conditions.  The two different approaches lead to the same on-shell action with an additional $\gamma$ term.

The purpose of this paper is to study the effect of this $\gamma$-term on the on-shell action.  For simple theory such as the Maxwell theory with ${\cal L}_F =- \fft14 F^2$, we have $\tilde F_{\mu\nu} = F_{\mu\nu}$ and hence
\be
{\cal I }^F_{\rm on-shell} =  \int_M d^4x \sqrt{-g} \Big(-\ft14 (1 - 2\gamma) F^2\Big) \,, \label{actFmax}
\ee
which vanishes when $\gamma=\ft12$.  Since the electric and magnetic contributions to the energy-momentum tensor and hence the metric are symmetric, it follows that the on-shell action with $\gamma=\ft12$ is symmetric with respect to the electric and magnetic charges.  However, for general ${\cal L}_{F}$, the quantity
${\cal I }^F_{\rm on-shell}$ does not necessary vanish even when $\gamma=\ft12$. This leads to a question whether it still has the electromagnetic duality.  We shall provide sufficiently enough examples that support the conjecture that $\gamma=\ft12$ is the universal coefficient for restoring the electromagnetic duality for the on-shell action, provided that the electromagnetic duality does exist in the equations of the theory.

In four dimensions, there are more total derivatives that one can consider.  One is the the $F\wedge F$ term and the other is the Gauss-Bonnet combination
\bea
I_{F\wedge F} &=&\fft{\kappa}{16 \pi} \, \tilde \gamma \int_M  d^4 x\, \fft 14 \epsilon ^{\mu\nu\rho\sigma} F_{\mu\nu} F_{\rho\sigma}\,,\nn\\
I_{\rm GB} &=& \alpha \int_M d^4 x \sqrt{-g} (R^2 - 4 R^{\mu\nu} R_{\mu\nu} + R^{\mu\nu\rho\sigma} R_{\mu\nu
\rho\sigma})\,.
\eea
Both terms are expected to have electromagnetic duality, in the type of minimally-coupled (generalized) Maxwell theories we consider in this paper.

Finally, it should be pointed out that equations (\ref{genbi}) and (\ref{tFeom}) interchange each other if we interchange ${*F}$ and $\widetilde F$, indicating the electromagnetic duality can always be implemented for the equations. However, the energy-momentum tensor expressed in $F$ and $\widetilde F$, namely $T_{\mu\nu}(F)$ and $T_{\mu\nu}(\widetilde F)$, may not have the same form and hence the electromagnetic duality can be broken by the Einstein equation.  In this paper, we shall only consider examples where the electromagnetic duality is indeed the symmetry for the equations and the metric of the dyonic black hole depends only on the combination $Q_e^2 + Q_m^2$.

\section{Einstein-Maxwell thoery}

In this section, we review the Einstein-Maxwell theory to illustrate a feature that the electric and magnetic contributions to the holographic complexity of a dyonic black hole are different and how this can be resolved \cite{goto}. Einstein-Maxwell theory in four dimensions, including the Gibbons-Howking term, is given by
\be
 I= \fft {\kappa}{16 \pi} \int_M d^4x \sqrt{-g} \big( R - 2 \Lambda  - \ft 14 F^2 \big)  + \fft{\kappa}{8 \pi} \int_{\partial M} d^3x \sqrt{-h} K \,, \label{actem}
\ee
where $h$ is the determinant of the boundary induced metric, $K$ is the trace of the extensive tensor and we set $\Lambda = - 3/\ell^2 $. The theory admits the dyonic Reissner-Nordstr\"om-AdS (RN-AdS) blakc hole.
\be
ds^2 = -f dt^2 + \fft{dr^2}{f} + r^2 (d\theta^2 + \sin^2\theta\, d\phi^2) \,, \quad  A= a dt + p \cos\theta d\phi\,,
\ee
with
\be
f = \fft{r^2}{\ell^2} +1 - \fft{\mu}{r} + \fft{q^2+p^2}{4 r^2} \,, \qquad a = \fft q r \,.
\ee
Here $(\mu, p, q)$ are integration constants. We choose the gauge that the electric potential $a$ vanishes at infinity. The mass of the black hole, electric charge, magnetic charge,  electric potential and magnetic potentials are
\be
M = \fft{\kappa \mu}{2} \,, \quad Q_e = \fft{\kappa q}{4} \,,\quad  Q_m = \fft{\kappa p}{4} \,, \quad \Phi_{e\pm} = \fft{q}{r_\pm }\,, \quad \Phi_{m\pm} = \fft{p}{r_\pm}\,,
\ee
where $\pm$ represent quantities evaluated on the outer and inner horizons of the black hole.

In order to calculate the late-time action growth, we follow the method in \cite{camyers} and we shall present only the main results. For more details we refer to \cite{camyers}. To apply this method, we introduce the null coordinates $(u,v)$, defined by
\be
du = dt + \fft{dr}{\sqrt {hf}} \,, \quad dv=dt - \fft{dr}{\sqrt{hf}}\,.
\ee
The Wheeler-De Witt patch of the black hole is shown in Fig.~1, which is surrounded by the light rays symmetrically anchored to the left and right boundary time slices with $t_L = t_R = t/2$. The action growth rate consists of three parts (bulk, boundary and joint),
\be
\fft{dI}{dt} = \fft{d I_{\rm bulk}}{dt} + \fft {d I_{\rm bd}}{dt} + \fft{d I_{\rm joint}}{dt} \,,
\ee
with
\bea
\fft{d I_{\rm bulk}}{dt} &=& \fft{\kappa}{4} \Big(  \frac{  \left(p^2-q^2\right)}{2 r}-2 g^2   r^3  \Big) \Big |_{r_-}^{r_+} \,, \nn\\
\fft {d}{dt}(I_{\rm joint}+I_{\rm bd}) &=& \fft{\kappa}{4} \Big(-\frac{ \left(p^2+q^2\right)}{2 r} + 2 g^2   r^3 \Big) \Big |_{r_-}^{r_+}  \,. \label{atmax}
\eea
Thus we see that the magnetic and electric contributions have opposite signs in the first term of the bulk contribution, originated from the kinetic $F^2$ term. There are cancelations among these parts, so that the total action growth rate is quite simple
\be
\fft{dI}{dt} = \fft{d I_{\rm bulk}}{dt} + \fft {d}{dt}(I_{\rm joint}+I_{\rm ct}) = - \fft{\kappa q^2}{4 r} \Big |_{r_-}^{r_+} = - \Phi_e Q_e \Big |_{r_-}^{r_+}  \,.
\ee
The result is independent of magnetic charge, in other words, the electric charge and the magnetic charge are not in the equal footing. In particular, the action vanishes for the purely magnetic black hole. It is worth commenting that this is a special result of a more general formula
\be
\fft{dI}{dt} = (F+ TS)_+ - (F +TS)_- = {\cal H}_+ - {\cal H}_-  \,,
\ee
shown in \cite{ptwoh} for all the black holes with both inner and outer horizons.

To address this issue, it was pointed out that at the price of introducing mixed boundary conditions in the variation principle, one can add a boundary term for the Maxwell field with an arbitrary coefficient $\gamma$ \cite{goto}
\be
I_{\mu Q} = \fft{\kappa \gamma}{16 \pi} \int_{\partial M} d \Sigma_\mu\, F^{\mu\nu} A_\nu \,,\qquad (i.e.\quad
\widetilde F^{\mu\nu}=F^{\mu\nu})\,. \label{maxwell}
\ee
On shell, with the help of the Maxwell equation and the Stokes' theorem, this term can be written as a bulk integration
\be
I_{\mu Q}\big |_{\rm on-shell} = \fft {\kappa \gamma}{32 \pi} \int_{M} d^4x \sqrt{-g} F^2 \,. \label{maxos}
\ee
Thus on shell, the total contributions of the Maxwell field to the action is modified to become
\be
I_{\rm Max} + I_{\mu Q} \big|_{\rm on-shell} = \fft{(2 \gamma -1) \kappa}{64 \pi} \int_M d^4x \sqrt{-g} F^2 \,.
\ee
Thus after including the boundary term, the total action growth rate in the WDW patch is now given by
\be
\fft{d I}{dt} = \Big(  -  (1-\gamma)\Phi_e Q_e -  \gamma \Phi_m Q_m  \Big)\Big |^{r_+}_{r_-}\,. \label{max4}
\ee
As can be expected, the action growth has electromagnetic duality when $\gamma=\ft12$.

At the end of section 2, two more total derivative terms were mentioned.  The Gauss-Bonnet term gives no contribution to the action growth for black holes with two horizons.  We find that the $F\wedge F$-term gives
\be
\fft{d I_{F\wedge F}}{dt} = -\frac{ \tilde \gamma \kappa p q}{2 r}\Big |_{r_-}^{r_+} \,.
\ee
It is clear that the action growth has electromagnetic duality. Compared to (\ref{max4}), it should perhaps be expressed as
\be
\fft{d I_{F\wedge F}}{dt} = -\tilde \gamma (\Phi_e Q_m + \Phi_m Q_e)\Big |_{r_-}^{r_+} \,.
\ee
Thus the most general form of the action growth rate that has the the electromagnetic duality is
\be
\fft{d I}{dt} = \Big(  - \ft12(\Phi_e Q_e +\Phi_m Q_m) -\tilde\gamma (\Phi_e Q_m + \Phi_m Q_e) \Big)\Big |^{r_+}_{r_-}\,,\label{max5}
\ee
where the coefficient $\tilde \gamma$ can be arbitrary.

\section{Einstein-Maxwell-Dilaton theory}

In \cite{goto}, the authors presented the complexity of a purely electric (or magnetic) AdS black hole in a class of EMD theories \cite{Gao:2004tu}. In this section, we go one step further and investigate AdS dyonic black hole solutions in certain EMD theories.   We consider
\be
{\cal L}=\sqrt{-g} \left(R - \ft12 (\partial\varphi)^2 - \ft14 e^{a\varphi} F^2 - V(\varphi)\right)\,,\qquad F=dA\,. \label{emd}
\ee
The scalar potential can be expressed in terms of a superpotential $W$ \cite{Lu:2013eoa}
\be
V = \Big(\fft{dW}{d\varphi}\Big)^2 - \fft34 W^2\,,\qquad
W=\fft{2\sqrt2\,g}{a^2+1}\,\big(e^{-\fft12 a\varphi} + a^2 e^{\fft1{2a}\varphi}\big)\,.
\ee
Explicitly, the scalar potential is \cite{Gao:2004tu}
\be
V = - \fft{2 g^2}{(1+a^2)^2} \Big[ a^2 (3 a^2 - 1) e^{\fft \varphi a} + (3 - a^2) e^{- a \varphi} + 8 a^2 e^{(-\fft a 2 + \fft{1}{2 a }) \varphi} \Big] \,.
\ee
Note that taking the Taylor series expansion in the region of $\varphi = 0$, we have
\be
V=-6g^2 - g^2 \varphi^2 + \fft{a^4-4a^2 +1}{24a^2} g^2 \varphi^4 -\fft{(a^2-\fft{1}{3})(a^2-1)(a^2-3)}{80a^3}
 g^2 \varphi^5 + \cdots\,.
\ee
Thus the limit of $a=0$ requires that $\varphi=0$. It follows from the kinetic term of the Maxwell field that the electromagnetic duality in the EMD theory
requires
\be
e^{a\varphi}{*F} \quad\leftrightarrow \quad F\,,\qquad \varphi\quad\leftrightarrow \quad-\varphi\,.
\ee
The scalar potential $V$ breaks this duality except when $a=1/\sqrt3$, 1 and $\sqrt3$, for which $V(\varphi)=V(-\varphi)$. Exact solutions of charged AdS dyonic black holes were known only for the latter two cases.  Thus we study the holographic complexity for $a=1$ and $a=\sqrt3$ dyonic black holes.

In order to recover the electromagnetic duality in the late-time action growth, a Maxwell boundary term is introduced \cite{goto}
\be
I_{\mu Q} = \fft{\kappa \gamma}{16 \pi} \int_{\partial M} d \Sigma_\mu \, e^{a \varphi} F^{\mu\nu} A_\nu \,,
\qquad (i.e.\quad \widetilde F^{\mu\nu} = e^{a \varphi} F^{\mu\nu}\,.)
\label{emdmaxbd}
\ee
As in the previous case, the on-shell action can be written as a bulk integratoin
\be
I_{\mu Q}\big |_{\rm on-shell} = \fft {\kappa \gamma}{32 \pi} \int_{M} d^4x \sqrt{-g} e^{-a\varphi} F^2\,.
\ee

\subsection{Dyonic black hole with $a = 1$}
When $a = 1$ the scalar potential can be written explicitly as
\be
V=-2g^2 (2 + \cosh\varphi)\,.
\ee
The dyonic black hole solution is \cite{Lu:2013eoa}
\bea
ds^2 &=& -(H_1 H_2)^{-1} f dt^2 + H_1 H_2 \Big(\fft{dr^2}{f} + r^2 (d\theta^2 + \sin^2\theta d\phi^2)\Big)\,,\cr
A&=&\sqrt{2} c_1 s_1^{-1} H_1^{-1} dt + \sqrt2 c_2 s_2 \cos\theta\,d\phi\,,\cr
\varphi &=& \log (H_1/H_2)\,,\qquad H_i=1 + \fft{\mu s_i^2}{r}\,,\qquad
f=1 - \fft{\mu}{r} + g^2 r^2 H_1^2H_2^2\,,
\eea
where $c_i=\cosh\delta_i$ and $s_i=\sinh\delta_i$. The mass, electric charge, electric potential, magnetic charge and potential are given by
\bea
M&=&\ft12 \mu (1 + s_1^2 + s_2^2)\,,\qquad Q_e=\ft14 \sqrt{2} \mu c_1 s_1\,,\qquad
Q_M=\ft14 \sqrt{2} \mu c_2 s_2\,, \cr
 \Phi_{e\pm} &=& \frac{\sqrt{2} c_1 \left(1-\frac{1}{H_1(r_\pm)}\right)}{s_1} \,, \qquad  \Phi_{m\pm} = \frac{\sqrt{2} c_2  \left(1-\frac{1}{H_2(r_\pm)}\right)}{s_2} \,,
\eea
where $r_\pm$ are the radius of the outer and inner horizon of the black hole.

Including the Maxwell boundary term (\ref{emdmaxbd}), the action growth rate can be written as
\be
\fft{dI}{dt} = \fft{d (I_{\rm bulk}+ I_{\mu Q})}{dt} + \fft {d I_{\rm bd}}{dt} + \fft{d I_{\rm joint}}{dt} \,,
\label{generalgrowth}
\ee
with
\bea
\fft{d (I_{\rm bulk} + I_{\mu Q})}{dt} &=& \Big(  \frac{ (r-\mu ) \left(r^2-\mu ^2 s_1^2 s_2^2\right)}{2\left(r+\mu  s_1^2\right) \left(r+\mu  s_2^2\right)} -  (1-\gamma) \Phi_e Q_e -  \gamma \Phi_m Q_m \Big) \Big|_{r_-}^{r_+} \,, \cr
 \fft{d (I_{\rm bd} + I_{\rm joint})}{dt} &=&  -  \frac{(r-\mu ) \left(r^2-\mu ^2 s_1^2 s_2^2\right)}{2\left(r+\mu  s_1^2\right) \left(r+\mu  s_2^2\right)}  \Big|_{r_-}^{r_+} \,.
\eea
The total action growth rate takes the identical form (\ref{max4}) as the Einstein-Maxwell theory.

\subsection{Dyonic black hole with $a = \sqrt 3$}

The case of  $a = \sqrt 3$ is the well-known Kuluza-Klein (KK) theory with the scalar potential
\be
V=-6g^2 \cosh(\ft1{\sqrt3}\varphi)\,.
\ee
The KK dyonic AdS black hole was constructed in \cite{Lu:2013ura}.  For simplicity, we consider only the planar geometry, for which the solution is
\bea
ds^2 &=& - (H_1H_2)^{-\fft12} f dt^2 + (H_1 H_2)^\fft12 \big( \fft{dr^2}{f} + r^2 (dx^2 + dy^2) \big) \,, \cr
\varphi &=& \fft{\sqrt 3}{2} \text{log} \fft{H_2}{H_1} \,, \qquad f = - \fft{2 \mu}{r} + g^2 r^2 H_1 H_2 \,, \cr
A &=& \sqrt{2 \mu} \big( \fft{r+ 2 \beta_1}{\sqrt{\beta_1} H_1 r} dt + 2 \sqrt{\beta_2} x dy \big) \,,\cr
H_1 &=& 1 + \fft{4 \beta_1}{r} + \fft{4 \beta_1 \beta_2}{r^2} \,, \qquad H_2 = 1 + \fft{4 \beta_2}{r} + \fft{4 \beta_1 \beta_2}{r^2} \,,
\eea
The solution has integration constants $(\mu, \beta_1, \beta_2)$, parameterizing the mass and electric and magnetic charges:
\be
M = \mu \,,\qquad Q_e = \sqrt{\fft{\mu \beta_1}{2}}\,,\qquad Q_m = \sqrt{\fft{\mu \beta_2}{2}} \,,
\ee
The electric and magnetic potentials on both inner and outer horizons $r_\pm$ are
\be
\Phi_{e\pm} = \fft{2 \sqrt{2 \mu \beta_1} ( r + 2 \beta_2 )}{r^2 H_1} \Big|_\pm \,, \qquad \Phi_{m\pm} = \fft{2 \sqrt{2 \mu \beta_2} ( r + 2 \beta_1 )}{r^2 H_2} \Big|_\pm \,,
\ee

Including the Maxwell boundary term (\ref{emdmaxbd}), the action growth rate is (\ref{generalgrowth}) and we find
\bea
\fft{d (I_{\rm bulk} + I_{\mu Q})}{dt} &=& \Big( -  (1-\gamma)\Phi_e Q_e -  \gamma \Phi_m Q_m   \cr
&&  -    \frac{ \mu  (2 \beta_1+r) (2 \beta_2+r) \left(r^2-4 \beta_1 \beta_2\right)}{\left(4 \beta_1 \beta_2+r^2+4 \beta_1 r\right) \left(4 \beta_1 \beta_2+r^2+4 \beta_2 r\right)} \Big) \Big|_{r_-}^{r_+} \,, \cr
 \fft{d (I_{\rm bd} + I_{\rm joint})}{dt} &=&   \frac{ \mu  (2 \beta_1+r) (2 \beta_2+r) \left(r^2-4 \beta_1 \beta_2\right)}{\left(4 \beta_1 \beta_2+r^2+4 \beta_1 r\right) \left(4 \beta_1 \beta_2+r^2+4 \beta_2 r\right)}  \Big|_{r_-}^{r_+} \,.
\eea
The total action growth rate is again (\ref{max4}).  Thus we see that the action growth rate of dyonic AdS black holes in EMD theories take the same form as that of Einstein-Maxwell theory, and it recovers the electromagnetic duality when $\gamma=\fft12$.

\section{Einstein-Born-Infeld theory}

The theory of Einstein-Born-Infeld (EBI) is a generalization of the Born-Infeld (BI) theory \cite{bi} to include gravity. The Lagrangian, with a bare cosmological constant $\Lambda_0$, is given by
\be
{\cal L} = \sqrt g ( R - 2 \Lambda_0  ) - b^2 \sqrt{ - \text{det} ( g_{\mu\nu} + \fft{F_{\mu\nu}}{b} ) } \,,
\ee
where $\Lambda_0 = \Lambda - \fft{b^2}{2}$ and $\Lambda$ is the effective cosmological constant.  In four dimensions, the theory admits AdS dyonic black hole solutions \cite{ebi}
\bea
ds^2 = - f dt^2 + \fft{dr^2}{f} + r^2 ( d\theta^2 + \sin^2\theta\, d\phi^2  ) \,, \qquad A = a dt + p \cos\theta\, d\phi \,,
\eea
with
\bea
f &=& - \fft13 \Lambda_0 r^2 + 1 - \fft{\mu}{r} - \fft{b^2}{6} \sqrt{ r^4 + \fft{Q^2}{b^2} } + \fft{Q^2}{3 r^2} {}_2F_1[ \ft14\,,\ft12\,;\ft54\,;- \ft{Q^2}{b^2 r^4} ] \,, \cr
a&=& \fft q r {}_2F_1[ \ft14\,,\ft12\,;\ft54\,;- \ft{Q^2}{b^2 r^4} ]\,, \qquad Q^2 = p^2 + q^2 \,,
\eea
where $(\mu,p,q)$ are the integration constants parameterizing the mass and charges of the black hole. To be specific, we have
\bea
M &=& \fft \mu 2 \,, \qquad Q_e =  \fft q 4 \,, \qquad Q_m = \fft p 4 \,, \nn\\
 \Phi_{e \pm}& =& \fft q r {}_2F_1[ \ft14\,,\ft12\,;\ft54\,;- \ft{Q^2}{b^2 r^4} ]\Big|_{r_\pm} \,, \qquad \Phi_{m \pm} = \fft p r {}_2F_1[ \ft14\,,\ft12\,;\ft54\,;- \ft{Q^2}{b^2 r^4} ]\Big|_{r_\pm} \,,
\eea
where $r_\pm$ are the radius of the outer and inner horizons. Thus we see that the metric, depending on $Q_e^2 + Q_m^2$,  has the electromagnetic duality.

Though the Maxwell field is singularity free, the black hole solution has a curvature singularity at $r=0$. When $\sqrt { p^2 + q^2} \le 2/b$, or  $\sqrt{p^2 + q^2} > 2/b$ but with $M\ge M^*$, the dyonic black hole has only one event horizon, where
\be
M^* = \fft{\Gamma(\fft 14)^2 \sqrt b}{24 \sqrt \pi } Q^{\fft 32}\,.
\ee
When $\sqrt{p^2 + q^2} > 2/b$ and $M_{\rm extrem}< M < M^*$, the solution has two horizons.  More details about the structure of the black hole singularity and the black hole thermodynamics can be found in  \cite{ebilu}.

We are now in the stage of calculating the on-shell action growth rate. As discussed in section 2, now the Maxwell boundary term is given by
\bea
I_{\mu Q} = \gamma \int_{\partial M} d\Sigma_\mu\, \widetilde F^{\mu\nu} A_\nu\,,\qquad
\widetilde F^{\mu\nu}= b \fft{\sqrt {-h}}{\sqrt{-g}} (h^{-1})^{[\mu\nu]} \,,
\eea
where
\be
h_{\mu\nu} = g_{\mu\nu} + \fft{F_{\mu\nu}}{b} \,, \qquad \sqrt h = \sqrt { - h_{\mu\nu}} \,, \qquad (h^{-1})^{[\mu\nu]} = \ft12 \big( (h^{-1})^{\mu\nu} - (h^{-1})^{\nu\mu} \big)\,,
\ee
and $(h^{-1})^{\mu\nu}$ is the inverse of $h_{\mu\nu}$. (There should be no confusion between the notation of this $h_{\mu\nu}$ and the boundary induced metric $h_{ab}$ in the Gibbons-Hawking surface term.)  The on-shell action can be expressed in terms of the bulk integration
\bea
I^{\mu Q}_{\rm on-shell}  = \int_M d^4x\,  \ft12 \gamma b  \sqrt h (h^{-1})^{[\mu\nu]} F_{\mu\nu} \,.
\eea
Having included this term, the total action growth rate is again given by (\ref{generalgrowth}), but
with
\bea
\fft{d (I_{\rm bulk} + I_{\mu Q})}{dt} &=& \Big( -  (1-\gamma) \Phi_e Q_e -  \gamma \Phi_m Q_m    \nn\\
&&  + \frac{1}{8} b  r \sqrt{p^2+q^2+b ^2 r^4}-\fft{ r}{4}+\fft{\Lambda}{4}  r^3 \Big) \Big|_{r_-}^{r_+} \,, \cr
 \fft{d (I_{\rm bd} + I_{\rm joint})}{dt} &=&  \Big( -\frac{1}{8} b  r \sqrt{p^2+q^2+b ^2 r^4}+\fft{ r}{4}-\fft{\Lambda}{4}  r^3 \Big)  \Big|_{r_-}^{r_+} \,.
\eea
Thus the later-time action growth rate again takes the same form as the Einstein-Maxwell case, given by (\ref{max4}).

For the dyonic black holes with only one horizon, the causal structure of the black hole is analogous to that of the Schwarzschild black hole. We follow the same procedure \cite{camyers} and find
\bea
\fft{dI}{d t} &=&  2 M  -  (1-\gamma) \Phi_{e+} Q_e -  \gamma \Phi_{m+} Q_m \, \nn\\
&& -\frac{5 \sqrt{\beta } \Gamma \left(\frac{1}{4}\right) \Gamma \left(\frac{5}{4}\right) \left(Q_e^2+Q_m^2\right){}^{3/4}}{3 \sqrt{\pi }}+ \frac{\sqrt{\beta } (1-2 \gamma_e) \Gamma \left(\frac{1}{4}\right) \Gamma \left(\frac{5}{4}\right) \left(Q_e^2-Q_m^2\right)}{\sqrt{\pi } \sqrt[4]{Q_e^2+Q_m^2}}\,. \label{ebi1h}
\eea
The result contains two parts. The first line of the above is related to the black hole event horizon and can be expressed in terms of thermodynamical quantities. The second line of formula comes from the contribution from the curvature singularity $r=0$. Both parts independently contain the electromagnetic duality when $\gamma=\fft12$.

The emergency of the electromagnetic duality at precise $\gamma=\fft12$ is much less trivial than the previous cases.  This is because now the duality-violating term (\ref{actF}) no longer simply vanish when $\gamma=\fft12$.  In fact, the on-shell Maxwell bulk contribution is given by
\bea
&&\fft{d}{dt}\Big(\fft{-b^2}{16 \pi} \int_M d^4x \sqrt{-\text{det} ( g_{\mu\nu} + \fft{F_{\mu\nu}}{b} )  }\Big) \cr
&=& \Big( \frac{1}{3} b  r \sqrt{16 (Q_e^2+ Q_m^2)+b ^2 r^4} -\frac{2 b  r    \left( Q_e^2 + Q_m^2 \right)  \, _2F_1[\frac{1}{4},\frac{1}{2};\frac{5}{4};-\frac{r^4 b ^2}{16 \left(Q_e^2+Q_m^2\right)}]}{3 \sqrt{Q_e^2+Q_m^2}}\nn\\
&&
+\frac{2 b  r  \left( Q_e^2- Q_m^2 \right) \big)\, _2F_1[\frac{1}{4},\frac{1}{2};\frac{5}{4};-\frac{r^4 b ^2}{16 \left(Q_e^2+Q_m^2\right)}]}{ \sqrt{Q_e^2+Q_m^2}}  \Big) \Big|^{r_+}_{r_-} \,,
\eea
while the Maxwell boundary term on-shell is given by
\be
\fft{d {\cal I}_{\mu Q}}{d t} = -\frac{4 b  \gamma_e r \left(Q_e^2-Q_m^2\right)  \, _2F_1[\frac{1}{4},\frac{1}{2};\frac{5}{4};-\frac{r^4 b ^2}{16 \left(Q_e^2+Q_m^2\right)}]}{\sqrt{Q_e^2+Q_m^2}} \Big|^{r_+}_{r_-} \,.
\ee
Together, the total Maxwell contribution is
\bea
\fft{d I_F}{d t}  &=& \Big( \frac{1}{3} b  r \sqrt{16 (Q_e^2+Q_m^2)+b ^2 r^4}
-\frac{2 b  r    \left( Q_e^2 + Q_m^2 \right)  \, _2F_1[\frac{1}{4},\frac{1}{2};\frac{5}{4};-\frac{r^4 b ^2}{16 \left(Q_e^2+Q_m^2\right)}]}{3 \sqrt{Q_e^2+Q_m^2}} \cr
&&+\frac{2 (1 - 2 \gamma_e) b  r  \left( Q_e^2- Q_m^2 \right) \big)\, _2F_1[\frac{1}{4},\frac{1}{2};\frac{5}{4};-\frac{r^4 b ^2}{16 \left(Q_e^2+Q_m^2\right)}]}{ \sqrt{Q_e^2+Q_m^2}}  \Big) \Big|^{r_+}_{r_-} \,.
\eea
We can see that when $\gamma = 1/2$, the Maxwell boundary term cancels the asymmetric part of the bulk contribution, which is proportional to $Q_e^2 - Q_m^2$. What is left is symmetric with $(Q_e,Q_m)$ that appears as $Q_e^2 + Q_m^2$ in the on-shell action.

\section{Einstein-Maxwell-Horndeski theory}

Einstein-Horndeski theory is a type of higher-derivative scalar-tensor theory where each field has no more than two derivatives in the equations of motion \cite{hd}. This feature is similar to that of Lovelock gravity \cite{ll}. A class of black hole solutions were constructed in Horndeski theory in \cite{hs1,hs2} and their thermodynamics were studied in \cite{th1,th2}, the instability and holographic applications can be found in \cite{stab1,stab2,stab3,holo1,holo2,holo3,holo4,hdap1,hdap2,hdap3,hdap4,hdap5,Li:2018rgn} and reference therein. In particular it was proposed that the theory is a holographic dual to some scale invariant but not conformal invariant quantum field theory \cite{hdap4,Li:2018rgn}. The theory we consider is given by
\be
{\cal S } = \fft{1}{16 \pi} \int_M \sqrt{-g} \Big[ \kappa (R - 2 \Lambda - \fft14 F^2 )   - \fft12 (\alpha g^{\mu\nu} - \eta G^{\mu\nu}) \partial_\mu \chi \partial_\nu \chi     \Big]\,,
\ee
where $G_{\mu\nu} = R_{\mu\nu} - \fft 12 R g_{\mu\nu}$ is the Einstein tensor, $\chi$ is the scalar field. It admits static AdS black hole solutions with both planar topology $\epsilon = 0$ and spherical topology $\epsilon = 1$\cite{hs2}.

\subsection{Charged AdS planar black hole}

The charged AdS planar black hole is given by
\bea
ds^2 &=& - h(r) dt^2 + \fft{dr^2}{f(r)} + r^2 (dx_1^2+ dx_2^2) \,,\nn\\
A &=& a(r) dt + \ft12 p(x_1 dx_2 - x_2 dx_1)\,,\qquad \chi = \chi(r) \,,
\eea
we find that the theory admits a black hole solution
\bea
f &=& \frac{36 g^4 r^8 (\beta  \eta +4 \kappa )^2}{\left(\kappa  (q^2+p^2)-6 g^2 r^4 (\beta  \eta +4 \kappa )\right)^2}\, h \,, \cr
h &=& g^2 r^2 -\frac{\mu}{r} + \frac{\kappa  (q^2+p^2)}{r^2 (\beta  \eta +4 \kappa )}-\frac{\kappa ^2 (q^2+p^2)^2}{60 g^2 r^6 (\beta  \eta +4 \kappa )^2} \,, \cr
\chi' &=& \sqrt { \beta -\fft{\kappa  (q^2+p^2)}{6 \eta g^2 r^4 } }\sqrt {\fft 1f } \,, \qquad
a = -\frac{q}{r} +\frac{\kappa  q (q^2+p^2)}{30 g^2 r^5 (\beta  \eta +4 \kappa )} \,,
\eea
with constraints
\be
\Lambda = -\frac{3 g^2 (\beta  \eta +2 \kappa )}{2 \kappa } \,, \quad \alpha = 3 \eta  g^2 \,.
\ee
The global structure and black hole thermodynamics was analysed in \cite{th2}.  Here we give some relevant thermodynamical quantities
\bea
M &=& \fft{4 \kappa + \beta \eta }{32 \pi} \, \mu  \,, \quad T =  -\frac{\kappa  \left(p^2+q^2\right)-6 g^2 r^4 (\beta  \eta +4 \kappa )}{8 \pi  r^3 (\beta  \eta +4 \kappa )}\big|_{r_0} \,, \cr
 \Phi_q &=&-\frac{q \left(\kappa  \left(p^2+q^2\right)-30 g^2 r^4 (\beta  \eta +4 \kappa )\right)}{30 g^2 r^5 (\beta  \eta +4 \kappa )} \big|_{r_0} \,, \quad Q_q = \fft{\kappa q}{16 \pi}\, , \cr
\Phi_p &=& -\frac{p \left(\kappa  \left(p^2+q^2\right)-30 g^2 r^4 (\beta  \eta +4 \kappa )\right)}{30 g^2 r^5 (\beta  \eta +4 \kappa )} \big|_{r_0} \,, \quad Q_p = \fft{\kappa p}{16 \pi}\, ,
\eea
Different from the RN-AdS black hole, there is an additional singularity $r_*$ where $f$ diverges
\be
\kappa ( p^2+q^2) -6 g^2 r_*^4 (\beta  \eta +4 \kappa)  = 0 \,.
\ee
There is a crucial difference with the RN black hole, this black hole has only one event horizon and thus has no extremal limit. Even though the temperature can be arbitrarily small, it can never reach zero \cite{th1,th2,feng}.

Now, we are in the position of calculating the action growth. Including the Maxwell boundary $\gamma$-term, we find that the total action growth is
\bea
\fft{dI}{dt} &=&  2 M  - (1-\gamma)\Phi_e Q_e - \gamma \Phi_m Q_m\nn\\
 &&-\frac{32 \pi  \left(Q_e^2+Q_m^2\right)}{3 \kappa  r_*} + \frac{32 \pi  (1-2 \gamma_e) \left(Q_e^2-Q_m^2\right)}{5 \kappa  r_*} \,.
\eea

\subsection{Charged AdS spherical black hole }

The solution is given by
\bea
ds^2 &=& - h(r) dt^2 + \fft{dr^2}{f(r)} + r^2 (d \theta^2 + \sin \theta^2 d \phi^2) \,,\nn\\
A &=& a(r) dt + p \, \cos \theta d\phi\,,\qquad \chi = \chi(r) \,,
\eea
with
\bea
f &=& \frac{4 r^4 \left(3 g^2 r^2+1\right)^2 (\beta  \eta +4 \kappa )^2}{\left(\kappa  \left(q^2 + p^2-8 r^2\right)-6 g^2 r^4 (\beta  \eta +4 \kappa )\right)^2}\, h \,, \cr
h &=& \frac{4 \kappa -\beta  \eta }{\beta  \eta +4 \kappa }+\frac{\kappa ^2 \left(p^2+q^2\right) \left(3 p^2 g^2+3 g^2 q^2+16\right)}{4 r^2 (\beta  \eta +4 \kappa )^2}-\frac{\kappa ^2 \left(p^2+q^2\right)^2}{12 r^4 (\beta  \eta +4 \kappa )^2}+g^2 r^2-\frac{\mu }{r} \cr
&&-\frac{\tan ^{-1}\left(\frac{1}{\sqrt{3} g r}\right) \left(2 \beta  \eta -3 g^2 \kappa  \left(p^2+q^2\right)\right)^2}{4 \sqrt{3} g r (\beta  \eta +4 \kappa )^2} \,, \cr
\chi' &=& \sqrt { \frac{6 \beta  \eta  g^2 r^4-\kappa ( q^2 + p^2 )}{2 \eta  r^2(1+3 g^2 r^2)} }\sqrt {\fft 1f } \,, \cr
a &=& -\frac{\kappa  q \left(3 g^2 (q^2+p^2)+8\right)}{ 2 r (\beta  \eta +4 \kappa )}+\frac{\kappa  q (q^2 + p^2)}{6 r^3  (\beta  \eta +4 \kappa )} \cr
&&+\frac{\sqrt{3} g q \tan ^{-1}\left(\fft{1}{\sqrt{3} g r}\right) \left(3 g^2 \kappa  (q^2+p^2)-2 \beta  \eta \right)}{2 (\beta  \eta +4 \kappa) } \,,
\eea
together with the constraints
\be
\Lambda = -\frac{3 g^2 (\beta  \eta +2 \kappa )}{2 \kappa } \,, \quad \alpha = 3 \eta  g^2 \,.
\ee
To be precise, the black hole is asymptotically locally AdS. The global structure and black hole thermodybamics was analysed in \cite{th2}. The relevant thermodynamical quantities are
\be
M = \fft{4 \kappa + \beta \eta }{8} \, \mu\,, \qquad Q_e = \fft{\kappa q}{4} \,, \qquad Q_m = \fft{\kappa p}{4} \,,\qquad \Phi_e = \fft{\xi q}{r_+}\,,\qquad \Phi_m=\fft{\xi p}{r_+}\,,
\ee
where
\be
\xi= \frac{\kappa  \left(24+(p^2+q^2)(9g^2-\fft{1}{r_+^2})\right)-3 \sqrt{3} g r_+ \tan ^{-1}\left(\frac{1}{\sqrt{3} g r_+}\right) \left(3 g^2 \kappa  \left(p^2+q^2\right)-2 \beta  \eta \right)}{6 (\beta  \eta +4 \kappa )}  \,,
\ee
The black hole has only one horizon and there is a curvature singularity at $r_*$ where $f$ diverges, namely
\be
\kappa (q^2 + p^2 - 8 r_*^2)-6 g^2 r_*^4 (\beta  \eta +4 \kappa )  = 0 \,.
\ee
We find that the total action growth (including the Maxwell boundary term) is given by
\bea
\fft{dI}{dt} &=&  2 M  - (1-\gamma) \Phi_e Q_e  - \gamma \Phi_m Q_m \cr
&& -\frac{1}{4} g^2 r_*^3 (\beta  \eta +4 \kappa )+r_* \left(\frac{\beta  \eta }{4}-\kappa \right) -\frac{8 \left(Q_e^2+Q_m^2\right)}{r_* (\beta  \eta +4 \kappa )} \cr
&& +\frac{\beta  \eta  \tan ^{-1}\left(\frac{1}{\sqrt{3} g r_*}\right) \left(\beta  \eta  \kappa -24 g^2 (Q_e^2+Q_m^2)\right)}{4 \sqrt{3} g \kappa  (\beta  \eta +4 \kappa )} \cr
&& + \frac{2 (1-2 \gamma) \left(Q_e^2-Q_m^2\right)}{3 \kappa ^2 r_*^3 (\beta  \eta +4 \kappa )} \Big(  3 \sqrt{3} g r_*^3 \tan ^{-1}\left(\frac{1}{\sqrt{3} g r_*}\right) \left(\beta  \eta  \kappa -24 g^2 Q_p^2-24 g^2 Q_q^2\right) \cr
&&  \qquad \qquad \qquad \qquad \qquad    +8 \left(9 g^2 r_*^2-1\right) ( Q_p^2 + Q_q^2 ) +12 \kappa ^2 r_*^2 \Big) \,.
\eea
Thus although the results appear to be more complicated, but it maintains the essential feature, namely the action growth rate preserves the electromagnetic duality when $\gamma=\fft12$.

\section{Conclusions}

   In this paper, we investigated the holographic complexity for dyonic black holes in Einstein-Maxwell-Dilaton
theory, Einstein-Born-Infeld theroy and Einstein-Maxwell-Horndeski theory by means of the CA conjecture. The theories all have the electromagnetic duality, but realized only at the level of equations. We present a universal $\gamma$-term in both first-order and second-order formalism so that the duality is restored in the on-shell action.  The advantage of the first-order formalism is that the variation principle involves only the standard Dirichlet boundary condition instead of the mixed boundary condition required by the second-order formalism.

For the dyonic black holes with either two horizons or just one horizon, we find that the late time action growth rate has the general form
\bea
\hbox{two horizons:}&&\qquad
\fft{d I}{dt} = \Big(  -  (1-\gamma)\Phi_e Q_e -  \gamma \Phi_m Q_m  \Big)\Big |^{r_+}_{r_-}\,;\nn\\
\hbox{one horizon:}&&\qquad \fft{dI}{d t} =  2 M  -  (1-\gamma) \Phi_{e+} Q_e -  \gamma \Phi_{m+} Q_m \cr
&&\qquad + U(Q_e^2 + Q_m^2) + (1-2\gamma) (Q_e^2 - Q_m^2) V(Q_e^2 + Q_m^2)\,,
\eea
where $U$ and $V$ functions depend on the detail of a specific theory. These example suggests that $\gamma=\fft12$ is universal for restoring the electromagnetic duality in the holographic complexity from the CA conjecture. The result can be easily established for theories involving only the linear Maxwell equations. But for general nonlinear field equations, an abstract proof is still lacking.  However, our explicit demonstration that $\gamma=\fft12$ holds for the EBI theory strongly suggests that it is indeed universal. It is of great interest to give a general proof or investigate further examples.

Our results do not address the question of which action equals the holographic complexity, raised in \cite{goto}.  For theories with electromagnetic duality, we may require $\gamma=\fft12$ to fix the action.  However, even in this case, we can add more terms such as $F\wedge F$ into the action that respect the duality.  For the RN-AdS black hole, the most general possibility of the holographic complexity based on CA that respects the electric-magnetic duality is given by (\ref{max5}) with arbitrary coefficient $\tilde \gamma$.  When the bulk theory itself does not have the electromagnetic duality, (e.g.~including $(F^2)^2$ term,) there is, {\it a priori}, no justification even to fix $\gamma=\fft12$ and it is of great interest to investigate whether $\gamma=\fft12$ is still special. Regardless the outcome, the CA conjecture requires some further refinements.

\section*{Acknowledgement}

H.-S.L.~is supported in part by NSFC (National Natural Science Foundation of China) Grants No.~11475148 and No.~11675144. H.L.~is supported in part by NSFC Grants No.~11875200 and No.~11475024.

\end{document}